\documentclass[journal,10pt]{IEEEtran}

\usepackage{cite}
\usepackage{graphicx}
\usepackage{amsmath}
\usepackage{array}
\usepackage{mdwmath}
\usepackage{amssymb}
\usepackage{mdwtab}
\usepackage{stfloats}
\usepackage[tight,footnotesize]{subfigure}
\usepackage{amsmath,amsthm}
\usepackage{threeparttable}
\usepackage{color}
\usepackage{url}
\usepackage{algpseudocode}
\usepackage{algorithm}
\usepackage{multicol}
\usepackage{amssymb}
\usepackage{epstopdf}

\algrenewcommand{\algorithmicrequire}{\textbf{Input:}}
\algrenewcommand{\algorithmicensure}{\textbf{Output:}}

\newtheorem{remark}{Remark}
\newtheorem{lemma}{Lemma}

\begin{document}
\title{ \LARGE Discrete RIS Enhanced Space Shift Keying MIMO System via Reflecting Beamforming Optimization}
\author{
Xusheng Zhu, Qingqing Wu, Wen Chen, Xinyuan He, Lexi Xu, and Yaxin Zhang
 \thanks{
 (\emph{Corresponding author: Qingqing Wu}.)}
\thanks{X. Zhu, Q. Wu, W. Chen, and X. He are with the Department of Electronic Engineering, Shanghai Jiao Tong University, Shanghai 200240, China (e-mail: xushengzhu@sjtu.edu.cn; qingqingwu@sjtu.edu.cn; wenchen@sjtu.edu.cn; euphoria\_0428@sjtu.edu.cn). L. Xu is with the Research Institute,
China United Network Communications Corporation, Beijing 100048,
China (e-mail: davidlexi@hotmail.com)
Y. Zhang is with the  University of Electronic Science and
Technology of China, Chengdu 610054, China (e-mail:
zhangyaxin@uestc.edu.cn).
}
}

\maketitle
\begin{abstract}
In this paper, a discrete reconfigurable intelligent surface (RIS)-assisted spatial shift keying (SSK) multiple-input multiple-output (MIMO) scheme is investigated, in which a direct link between the transmitter and the receiver is considered.
To improve the reliability of the RIS-SSK-MIMO scheme, we formulate an objective function based on minimizing the average bit error probability (ABEP). Since the reflecting phase shift of RIS is discrete, it is difficult to address this problem directly. To this end, we optimize the RIS phase shift to maximize the Euclidean distance between the minimum constellations by applying the successive convex approximation (SCA) and penalty-alternating optimization method.
Simulation results verify the superiority of the proposed RIS-SSK-MIMO scheme and demonstrate the impact of the number of RIS elements, the number of phase quantization bits, and the number of receive and transmit antennas in terms of reliability.
\end{abstract}

\section{Introduction}

To meet the growing communication demands, conventional multiple-input multiple-output (MIMO) technique is adopted, which can improve system effectiveness and reliability by utilizing space-time resources for multiplexing and diversity \cite{wang2024a}. However, MIMO has drawbacks like high complexity, cost and power consumption\cite{zhu2024per}. In response, reconfigurable intelligent surface (RIS), a planar structure of numerous low-cost and low-power reflective elements that can control wireless signal propagation by adjusting their phase and amplitude, has attracted wide attention for its unique advantages such as enhancing spectrum utilization efficiency, reducing power consumption and offering high flexibility \cite{wu2014int}.
Meanwhile, spatial modulation (SM) \cite{zhu2022on,zhu2024onx}, which uses transmit antenna indices to transmit information and improve spectral efficiency and reduce system complexity, and spatial shift keying (SSK), which transmits information only by activating different transmit antenna positions and provides new ideas for wireless communication development, are other solutions to address insufficient spectrum\cite{zhu2024ris}.

Inspired by the advantages of RIS, \cite{can2022on} investigated the RIS-assisted SSK system. It presents a mathematical framework and provides simulation results, indicating that the scheme enables highly reliable transmission with high energy efficiency.
Besides, \cite{singh2022ris} studied RIS-assisted SSK modulation and reflection phase modulation scheme where RIS embeds information in  reflection phase shift.
To improve the spectral efficiency, \cite{zhu2023ris} and \cite{zhu2024robutst} studied the full-duplex SSK system based on RIS, and investigated the cases of perfect and imperfect channel state information (CSI), respectively.
It is worth noting that the RIS-assisted SSK technique in \cite{can2022on,singh2022ris,zhu2023ris,zhu2024robutst} is implemented at the transmit end.
In contrast, the RIS-assisted SSK technology in \cite{zhang2023received,dian2022ris,yuan2021rec,marin2024ris} is implemented at the receive end.
To be specific, \cite{zhang2023received} designed a RIS-assisted generalized space-shift keying (GSSK) based antennas selection scheme where GSSK works at receiver and RIS elements are adjusted for antennas selection to maximize signal-to-noise ratio (SNR).
Moreover, \cite{yuan2021rec} proposed two RIS-based SSK schemes, RIS-SSK passive beamforming which employs RIS for beamforming to maximize minimum squared Euclidean distance and RIS-SSK  Alamouti space-time block coding which employs RIS for ASTBC and transmits its own Alamouti-coded information while reflecting SSK signals.
Additionally, \cite{marin2024ris} enhances wireless communication systems with RIS by introducing RIS-assisted receive GSSK.
Furthermore,
\cite{dian2022ris} investigated RIS-receive quadrature SSK to enhance spectral efficiency via independent use of real and imaginary dimensions. By using low-complexity greedy detector, a max-min optimization for SNR is performed.
The above work is to align all the reflected phase shifts of RIS with a single target antenna, obtain the composite channel distribution, and then derive the closed-form ABEP. However, in the RIS-assisted SSK-MIMO system, RIS cannot eliminate the phase of multiple antennas in the receive array, so the previous method is no longer applicable.

Against this background, we study a RIS-SSK-MIMO system model that considers discrete RIS phase shift. Then, we formulate an optimization problem to minimize the average bit error probability (ABEP) by designing the reflecting coefficients of the RIS.
To simplify the problem, we convert it into a problem of maximizing the minimum constellation point Euclidean distance subject to the discrete phase shift of RIS.
To tackle this issue, we adopt successive convex approximation (SCA) and penalty-alternative method to get a suboptimal solution. Lastly, simulation results show that the proposed RIS-SSK-MIMO outperforms the benchmarks in term of ABEP.

\section{System Model}
As shown in Fig. 1, we consider a RIS-assisted MIMO system, where  the transmitter (Tx) and receiver (Rx) are equipped with $N_t$ and $N_r$ antennas, respectively.  The RIS consists of $L$ passive reflecting units, and each element can reflect the incident signal independently.
To characterize the limits of RIS, we set the reflection amplitude of each element to one.
In this manner, the vector $\mathbf{v}=[e^{j\theta_1},e^{j\theta_2},\cdots,e^{j\theta_l},\cdots,e^{j\theta_L}]^T$ is used to denote the coefficients of the RIS, where $\theta_l$ stands for the reflection phase shift of $l$-th element of the RIS.
Due to hardware constraints, each reflecting unit can only select a finite number of discrete values in realistic cases. Without loss of generality, let $Q$ denote the number of bits of phase shift control for each RIS unit, thus we have
\begin{equation}
\mathbf{v}[n]\in\Phi=
\{
e^{j\theta_n}|\theta_n\in\Theta
\}, \forall n,
\end{equation}
where $\Theta=\{0,\frac{2\pi}{2^{Q}},\cdots,\frac{2\pi(2^{Q-1})}{2^{Q}}\}$, i.e., the discrete phase-shift values are assumed to be equally spaced in the interval $[0,2\pi)$.
If the $Q\to\infty$, the phase shift of RIS becomes the continuous phase shift control.

In the RIS-SSK-MIMO system, the arriving Rx signal consists of the direct link and the link which is reflected by the RIS.
To be specific, the channel between the Tx and Rx, Tx and RIS, RIS and Rx are $\mathbf{H}=L_{\rm Tx-Rx}\mathbf{H_0}\in\mathbb{C}^{N_r\times N_t}$,
$\mathbf{F}=L_{\rm Tx-RIS}\mathbf{F}_0\in\mathbb{C}^{L\times N_t}$, and $\mathbf{G}= L_{\rm RIS-Rx}\mathbf{G}_0\in\mathbb{C}^{N_r\times L}$, respectively, where $L_{\rm UE-Tx}$, $L_{\rm Tx-RIS}$, and $L_{\rm RIS-Rx}$ denote the corresponding path loss, and $\mathbf{H}_0$, $\mathbf{F}_0$, $\mathbf{G}_0$ denote the small scale fading, respectively.

In this paper, we assume that the path fading from Tx to Rx is relatively large. To enhance the received signal, we deploy RIS in the channel to construct line-of-sight (LoS) from Tx-RIS and RIS-Rx links, respectively.
As such, we model the $\mathbf{H}_0$ as a Rayleigh fading channel, and $\mathbf{G}_0$ and $\mathbf{F}_0$ as Rician fading channel.

\begin{figure}[t]
  \centering
  \includegraphics[width=6cm]{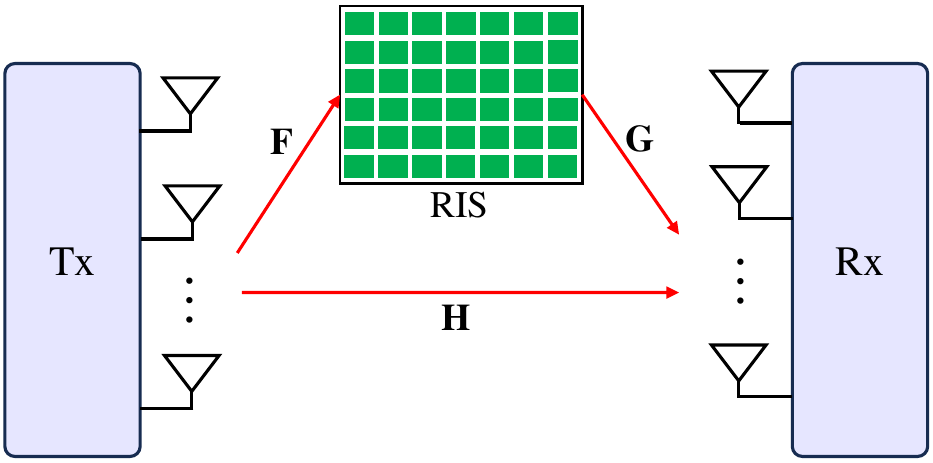}\\
  \caption{\small System model.}\label{sysmodel}
\end{figure}
\subsection{Transmission}
In SSK technique, the symbol bit does not transmit information. Specifically, there is usually only one RF chain available at the Tx, which means that only one antenna can be activated for each symbol transmission. To fully utilize the degrees of freedom provided by multiple antennas, the essence of SSK technology is to use the index of transmit antennas to convey information.
During each time slot, the bits used for transmitting information are reassigned.
Since the symbol bit does not transmit information, these bits are used to select spatial constellation points, which are elements of set
\begin{equation}
\mathcal{S}=\{\mathbf{e}_1,\mathbf{e}_2,\cdots,\mathbf{e}_{n_t},\cdots,\mathbf{e}_{N_t}\},
\end{equation}
where $\mathbf{e}_{n_t}$ denotes the $n_t$-th column of $N_r\times N_t$ identity matrix, which means the $n_t$-th antenna of Tx is activated.
As a result, the SSK symbol is given by $\mathbf{x}=\mathbf{e}_{n_t}\in\mathcal{S}$.

For a particular transmitted SSK symbol $\mathbf{x}$, the received signal at the Rx is as
\begin{equation}
\mathbf{y} = \mathbf{(H+GVF)x} + \mathbf{n},
\end{equation}
where $\mathbf{V}$ is a diagonal matrix ${\rm diag}\{\mathbf{v}\}$ consisting of RIS reflection coefficients, $\mathbf{n}\sim\mathcal{CN}(0,N_0\mathbf{I}_{N_r})$ represents the additive white Gaussian noise (AWGN) at the Rx side.


At the Rx side, the maximum likelihood (ML) algorithm is used to recover the original transmit signal as
\begin{equation}\label{ml1}
    {\hat{\mathbf{x}}} = \arg\mathop{\min}_{\mathbf{x}\in\mathcal{S}}\|\mathbf{y}-\mathbf{(H+GVF)x}\|^2,
\end{equation}
where $\mathbf{\hat x}$ stands for the detected signal $\mathbf{\hat x}=\mathbf{e}_{\hat n_t}\in\mathcal{S}$.
\subsection{ABEP}
Based on (\ref{ml1}), the CPEP that the Rx erroneously detects can be given by
\begin{equation}\label{cpep1}
\begin{aligned}
P_b=&\Pr\left(
\|\mathbf{y}-\mathbf{(H+GVF)x}\|^2>\|\mathbf{y}-\mathbf{(H+GVF)\hat x}\|^2
\right)\\
=&\Pr\left(
\|\mathbf{n}\|^2>\|\mathbf{n}-\mathbf{(H+GVF)(x-\hat x)}\|^2
\right)\\
=&\Pr\left(
2\Re\{\mathbf{n(H+GVF)(x-\hat x)}\}\right.\\&\left.-\|\mathbf{(H+GVF)(x-\hat x)}\|^2>0
\right)\\
=&\Pr(T>0),
\end{aligned}
\end{equation}
where the variable $T=2\Re\{\mathbf{n(H+GVF)(x-\hat x)}\}-\|\mathbf{(H+GVF)(x-\hat x)}\|^2$ follows the real Gaussian distribution, i.e., $T\sim\mathcal{N}(\mu_T,\sigma_T^2)$, where $\mu_T = -\|\mathbf{(H+GVF)(x-\hat x)}\|^2$ and $\sigma_T^2 = 2N_0\|\mathbf{(H+GVF)(x-\hat x)}\|^2$.
Thus, the (\ref{cpep1}) can be calculated as
\begin{equation}\label{cpep2}
P_b=Q\left(\sqrt{{\mu_T^2}/{\sigma_T^2}}\right)=Q\left(\sqrt{\frac{\Delta(\mathbf{v})}{2N_0}}\right),
\end{equation}
where $Q(\cdot)$ is the Gaussian right tail probability function, and $\Delta(\mathbf{v})= \|\mathbf{(H+GVF)(x-\hat x)}\|^2$.
In this case, the union upper bound of ABEP can be given as
\begin{equation}
{\rm ABEP} \leq \sum_{n_t=1}^{N_t}\sum_{\hat n_t = 1}^{N_t}\frac{P_bN(n_t\to\hat n_t)}{N_t\log_2N_t},
\end{equation}
where $N(n_t\to\hat n_t)$ denotes the Hamming distance between the transmit signal and detected signal.

\begin{remark}
Since the signal is received by multiple antennas simultaneously after being reflected by the RIS.
Meanwhile, the distribution of the numerator term becomes unpredictable after the signal has been conditioned by the RIS reflections, thus we cannot obtain the closed-form expression for (\ref{cpep2}) by adopting the \cite{zhu2024ris} method.
\end{remark}

\section{Proposed Solution}

\subsection{Problem Formulation}
The aim of (\ref{cpep2}) is to design the reflecting coefficients $\mathbf{v}$ of the RIS to minimize the ABEP.
Due to the difficulty in obtaining the closed-form expression of ABEP in the high SNR region.
In consideration of the fact that ABEP is dominated by received signal pairs of minimum Euclidean distance, we adopt the minimum mean Euclidean distance criterion for signal design.
Accordingly, we have the following optimization problem
\begin{equation}\label{formulateo}
\begin{aligned}
& \max _{\mathbf{v}} \min _{\forall n_t \neq\hat n_t}\left\{\Delta(\mathbf{v})\right\} \\
& \text { s.t. }   \mathbf{v}[l] \in \Phi, \quad l=1,2, \cdots, L.
\end{aligned}
\end{equation}

Problem (\ref{formulateo}) is a non-convex problem since $\mathbf{v}[n]$ has discrete property.
In general, it is difficult to find the optimal solution for this problem.
To deal with the discrete phase, we introduce the auxiliary vector $\mathbf{u}$.
Let us define $\mathbf{v=u}$, then the problem of (\ref{formulateo}) can be reformulated as
\begin{equation}\label{formulate1}
\begin{aligned}
& \max _{\mathbf{v}} \min _{\forall n_t \neq\hat n_t}\left\{\Delta(\mathbf{v})\right\} \\
& \text {s.t.} \quad \mathbf{v=u},\\
& \quad\quad  \mathbf{u}[l] \in \Phi, \quad l=1,2, \cdots, L.
\end{aligned}
\end{equation}

In this step, we resort to with the penalty function to remove the equality constrain $\mathbf{v=u}$. To be specific,
Concretely, we transform the equation constraints into penalty terms to yield
\begin{equation}\label{formulate2}
\begin{aligned}
& \max _{\mathbf{v}} \min _{\forall n_t \neq\hat n_t}\left\{\Delta(\mathbf{v})\right\}-\rho\|\mathbf{v-u}\|^2 \\
& \text {s.t.} \quad  \mathbf{u}[l] \in \Phi, \quad l=1,2, \cdots, L,
\end{aligned}
\end{equation}
where $\rho$ is a positive coefficient that penalizes violations of equality constraint.
In each iteration, the penalty coefficient becomes larger. Specifically, at the beginning of the iteration, $\rho$ is a very small value, and the purpose of this is to get a good starting point, even if this point is infeasible. As the number of iterations increases, the value of $\rho$ continues to increase, that is, $\rho=C\rho$, where $C$ is a positive number greater than 1. To this end, we can maximize the objective function and obtain a solution that satisfies the equality constraints in (\ref{formulate1}).

For each specific $\rho$ value, it remains difficult to address problem (\ref{formulate2}) directly. For this reason, we decompose (\ref{formulate2}) into two subproblems and employ alternating iterations to tackle them. Specifically, in the first subproblem, we optimize the reflection coefficient $\mathbf{v}$ by fixing $\mathbf{u}$. While in the second problem, we optimize $\mathbf u$ based on the $\mathbf{v}$ obtained from the optimization in the first problem.

\subsection{Proposed Iterative Design Method for $\mathbf{v}$}
For a given $\mathbf{u}$, we design $\mathbf{v}$. At this time, the optimization problem for $\mathbf{v}$ can be characterized as
\begin{equation}\label{optsk1}
\begin{aligned}
&\max_{\mathbf{v}}\mathop{\min}_{\forall n_t\neq \hat n_t}\left\{\Delta(\mathbf{v})\right\}-\rho\|\mathbf{v}-\mathbf{u}\|^2.
\end{aligned}
\end{equation}
For the $\Delta(\mathbf{v})$ term of (\ref{optsk1}), we need to reform it into a more tractable form as
\begin{equation}\label{optsk2}
\begin{aligned}
\Delta(\mathbf{v})
&= \|\mathbf{(H+GVF)}(\mathbf{x}-\mathbf{\hat x})\|^2\\
&= \|(\mathbf{h}_{n_t}+\mathbf{GVf}_{n_t})-(\mathbf{h}_{\bar k}+\mathbf{GVf}_{\hat n_t})\|^2\\
&=\|(\mathbf{h}_{n_t}+\mathbf{GD}_{n_t}\mathbf{v})-(\mathbf{h}_{\hat n_t}+\mathbf{GD}_{\hat n_t}\mathbf{v})\|^2,
\end{aligned}
\end{equation}
where $\mathbf{D}_{n_t} ={\rm diag}\{\mathbf{f}_{n_t}\}$ denotes the diagonal matrix, $\mathbf{h}_{n_t}$ and $\mathbf{f}_{n_t}$ stand for the $n_t$-th column of $\mathbf{H}$ and $\mathbf{F}$, respectively.
Without loss of generality, we define $\mathbf{a}=\mathbf{h}_{n_t}-\mathbf{h}_{\hat n_t}$ and
$\mathbf{B}=\mathbf{G}(\mathbf{D}_{n_t}-\mathbf{D}_{\hat n_t})$. Based on this, (\ref{optsk2}) can be given as
\begin{equation}
\begin{aligned}
\Delta(\mathbf{v})
&=\|\mathbf{a}+\mathbf{B}\mathbf{v}\|^2=\mathbf{v}^H\mathbf{R}\mathbf{v}+2\Re\{\mathbf{c}\mathbf{v}\}+m,
\end{aligned}
\end{equation}
where $\mathbf{R}=\mathbf{B}^H\mathbf{B}$, $\mathbf{c}=\mathbf{a}^H\mathbf{B}$, $m =\|\mathbf{a}\|^2$, and $\Re\{\cdot\}$ denotes the real part of a complex number.

In this manner, we can recast the (\ref{optsk1}) as
\begin{equation}\label{optsk3}
\begin{aligned}
&\max_{\mathbf{v}}\mathop{\min}_{\forall n_t\neq \hat n_t}
f(\mathbf{v})
,
\end{aligned}
\end{equation}
where $f(\mathbf{v}) = \mathbf{v}^H\mathbf{R}\mathbf{v}+2\Re\{\mathbf{c}\mathbf{v}\}+m-\rho\|\mathbf{v-u}\|^2$.
Here, (\ref{optsk3}) is still a non-convex problem since $f(\mathbf{v})$ is non-convex.
In spite of this, $\mathbf{v}^H\mathbf{R}\mathbf{v}$ and $\|\mathbf{v-u}\|^2$ are both convex functions with respect to $\mathbf{v}$.
As such, $f(\mathbf{v})$ has a difference-of-convex structure.
To tackle this issue, we can employ the SCA method to approximate it iteratively.
To be specific, the convex function $\mathbf{v}^H\mathbf{R}\mathbf{v}$ is approximated by employing first-order Taylor expansion at a given feasible point $\mathbf{v}_n$ as
\begin{equation}
\mathbf{v}^H\mathbf{R}\mathbf{v}\geq
2\Re\{\mathbf{v}_n^H\mathbf{R}\mathbf{v}\}-\mathbf{v}_n^H\mathbf{R}\mathbf{v}_n.
\end{equation}
Consequently, we can approximate $f(\mathbf{v})$ as
\begin{equation}
f(\mathbf{v})\geq -\rho\|\mathbf{v-u}\|^2+2\Re\{\mathbf{c}_n\mathbf{v}\}+m_{n},
\end{equation}
where $\mathbf{c}_n=\mathbf{c}+\mathbf{v}_n^H\mathbf{R}$ and $m_n=m-\mathbf{v}_n^H\mathbf{R}\mathbf{v}_n$.
In this regard, we can approximate the problem (\ref{optsk3}) as
as
\begin{equation}\label{optsk4}
\begin{aligned}
&\max_{\mathbf{v}}\mathop{\min}_{\forall n_t\neq \hat n_t}
-\rho\|\mathbf{v-u}\|^2+2\Re\{\mathbf{c}_{n}\mathbf{v}\}+m_{n}.
\end{aligned}
\end{equation}
It can be observed from (\ref{optsk4}) that the term of $-\rho\|\mathbf{v-u}\|^2$ is a concave function with respect to $\mathbf{v}$, and $2\Re\{\mathbf{c}_{n}\mathbf{v}\}$ is an affine function with respect to $\mathbf{v}$.
As such, the object function is to find the minimum of several concave functions, which is a concave function.
Thus, (\ref{optsk4}) is the convex function can be addressed efficiently.
To solve this problem, (\ref{optsk4}) can iteratively solve at different feasible points until convergence.
Let us denote the optimal solution  of (\ref{optsk4}) at the $n$-th iteration as $\mathbf{v}_n$.
Then, the value of $\mathbf{v}_n$ is set as the initial point of (\ref{optsk4}) in the $(n+1)$-th iteration.
This loop iterates until $\|\mathbf{v}_{n+1}-\mathbf{v}_{n}\|^2\leq \epsilon$ stops, where $\epsilon$ is the convergence threshold.
The detail iterations is summarized from line 4 to line 8 in {\bf Algorithm 1}.

\subsection{Proposed Design Method for $\mathbf{u}$}
After addressing problem (\ref{optsk3}), we fix $\mathbf{v}$ as a valid solution and optimize $\mathbf{u}$ by forming the following problem.
\begin{equation}\label{prou1}
\begin{aligned}
&\min_{\mathbf{u}}\|\mathbf{v-u}\|^2\\
&{\rm s.t.} \quad \mathbf{u}[l]\in\Phi, l= 1,2,\cdots,L.
\end{aligned}
\end{equation}

For this problem, $\mathbf{u}_l$ is decoupled in both the objective function and the constraints. Hence, the solution of (\ref{prou1}) can be obtained as
\begin{equation}\label{phas1}
\mathbf{u}[l] = e^{j\hat{\theta}_l},
\end{equation}
where we have
$
\hat{\theta}_l = \mathop{\arg\min}\limits_{\theta_l\in\Theta}\left|
\theta_l-\angle\mathbf{v}[l]
\right|,
$
where $\angle(\cdot)$ denotes the corresponding phase.

\subsection{Alternative Opimization for $\mathbf{v}$ and $\mathbf{u}$}
For a given value of $\rho$, the problem X can be tackled by alternating optimization of $\mathbf{v}$ and $\mathbf{u}$. It is worth noting that from line 4 to line 10 in {\bf Algorithm 1}, we describe in detail the alternating optimization process for $\mathbf{v}$ and $\mathbf{u}$.

\begin{algorithm}[t]
\caption{Address (\ref{formulate1}) via Penalty-Alterative Algorithm}\label{alg3}
\begin{algorithmic}[1]
\Require $\zeta \leftarrow 0$, $\mathbf{v}_\zeta$, $\mathbf{u}_\zeta$, $\rho$, $C$, $\epsilon$.
\Ensure $\mathbf{v}\leftarrow\mathbf{u}_{\zeta+1}$.\\
{\bf repeat}
\State\quad $\eta \leftarrow 0$, $\mathbf{v}_\eta\leftarrow\mathbf{v}_\zeta$, $\mathbf{u}_\eta\leftarrow\mathbf{u}_\zeta$.
\State \quad {\bf repeat}
\State\quad\quad $n \leftarrow 0$, $\mathbf{v}_n\leftarrow\mathbf{v}_\eta$, $\mathbf{u}_n\leftarrow\mathbf{u}_\eta$.
\State \quad\quad {\bf repeat}
\State \quad\quad\quad By resolving (\ref{optsk4}), we obtain $\mathbf{v}_{n+1}$.
\State \quad\quad\quad Update $n \leftarrow n+1$.
\State \quad\quad {\bf until}  $\|\mathbf{v}_{n+1}-\mathbf{v}_{n}\|^2\leq \epsilon$.
\State \quad\quad Update $\mathbf{u}_{\eta+1}$ via (\ref{phas1}).
\State\quad\quad Update $\eta \leftarrow \eta+1 $
\State \quad {\bf until}  $\|\mathbf{v}_{\eta+1}-\mathbf{v}_{\eta}\|^2\leq \epsilon$.
\State\quad Update $\rho \leftarrow C\rho $.
\State\quad Update $\zeta \leftarrow \zeta+1 $. \\
{\bf until}  $\|\mathbf{v}_{\zeta+1}-\mathbf{u}_{\zeta+1}\|^2\leq \epsilon$.
\end{algorithmic}
\end{algorithm}

\begin{lemma}
For given $\rho$, the alternating iteration optimization method  is converge.
\end{lemma}

\emph{Proof:} We let $Q(\mathbf{v}_\eta,\mathbf{u}_\eta)$ stand for the target value of problem (\ref{formulate2}) in the $\eta$-th iteration in {Algorithm 1}.
Besides, $Q_1(\mathbf{v}_\eta,\mathbf{u}_\eta)$ and $Q_2(\mathbf{v}_\eta,\mathbf{u}_\eta)$ denote the target values of (\ref{optsk3}) and (\ref{prou1}) in the $\eta$-th iteration, respectively.
Herein, we have
\begin{small}
\begin{equation}\label{conver1}
Q(\mathbf{v}_{\eta},\mathbf{u}_{\eta})\overset{(a)}{=}Q_1(\mathbf{v}_{\eta},\mathbf{u}_{\eta})\overset{(b)}{\leq}Q_1(\mathbf{v}_{\eta+1},\mathbf{u}_{\eta})\overset{(c)}{=}Q(\mathbf{v}_{\eta+1},\mathbf{u}_\eta),
\end{equation}
\end{small}%
where (a) and (c) satisfy the definitions of $Q(\mathbf{v}_{\eta},\mathbf{u}_{\eta})$ and $Q_1(\mathbf{v}_{\eta},\mathbf{u}_{\eta})$, $(b)$ denotes $\mathbf{v}_{\eta+1}$ is obtained with the SCA method, which is the max-min method and the objective value of (\ref{optsk3}) increases in each iteration.
The (\ref{conver1}) shows that the target function value of problem (\ref{formulate2}) is non-decreasing at each update of $\mathbf{v}$.
For the updating $\mathbf{u}$, we have
\begin{equation}\label{conver2}
\begin{aligned}
Q(\mathbf{v}_{\eta+1},\mathbf{u}_\eta)&\overset{(a)}{=}Q_2(\mathbf{v}_{\eta+1},\mathbf{u}_\eta)\overset{(d)}{\leq}
Q_2(\mathbf{v}_{\eta+1},\mathbf{u}_{\eta+1})\\
&\overset{(c)}{=}Q(\mathbf{v}_{\eta+1},\mathbf{u}_{\eta+1}),
\end{aligned}
\end{equation}
where $(d)$ means that each update of $\mathbf{u
}$ in (\ref{prou1}) is from the direction of reducing the value of the objective function.

According to (\ref{conver1}) and (\ref{conver2}), we can obtain $Q(\mathbf{v}_{\eta},\mathbf{u}_{\eta})\leq Q(\mathbf{v}_{\eta+1},\mathbf{u}_{\eta+1})$, which means the optimization problem is non-decreasing in each iteration.
Since (\ref{formulate2}) is a continuous function on a finite feasible set, its upper limit is finite positive number, so the optimization problem is guaranteed to converge.
$\hfill\blacksquare$

\subsection{Penalty-Alterative Algorithm}
The optimization problem (\ref{formulate1}) can be addressed with three layers of iterations. At the first layer of iterations, the value of $\rho$ is gradually increased at each iteration to force $\mathbf{v=u}$. In the second layer of iterations, $\mathbf{v}$ and $\mathbf{u}$ are optimized alternately. In the third layer iteration, the SCA method is adopted to find the solution (\ref{optsk3}), respectively. The whole solution process is shown in {\bf Algorithm 1}.

\subsection{Complexity Analysis}
For a specific $\rho$, $\mathbf{v}$ and $\mathbf{u}$ are optimized alternatively, where the main source of complexity is the solution process at each iteration. Specifically, the SCA method is used to solve problem (\ref{optsk1}), which requires successive quadratic problems with a complexity of $\mathcal{O}(L^3 + L^2(L + N_t))$, where $L$ is the dimensionality of the problem, and $L+N_t$ is the number of constraints. The complexity of (\ref{prou1}) is $\mathcal{O}(2^Q)$. Consequently, the total complexity is $\mathcal{ O}(L_1L_2(L_3(L^3 + L^2(L + M))) + 2^Q))$, where $L_1$ is the number of iterations at the first layer, $L_2$ is the number of iterations at the second layer, and $L_3$ is the number of iterations required by the SCA method.

\begin{figure}[t]
  \centering
  \includegraphics[width=4.2cm]{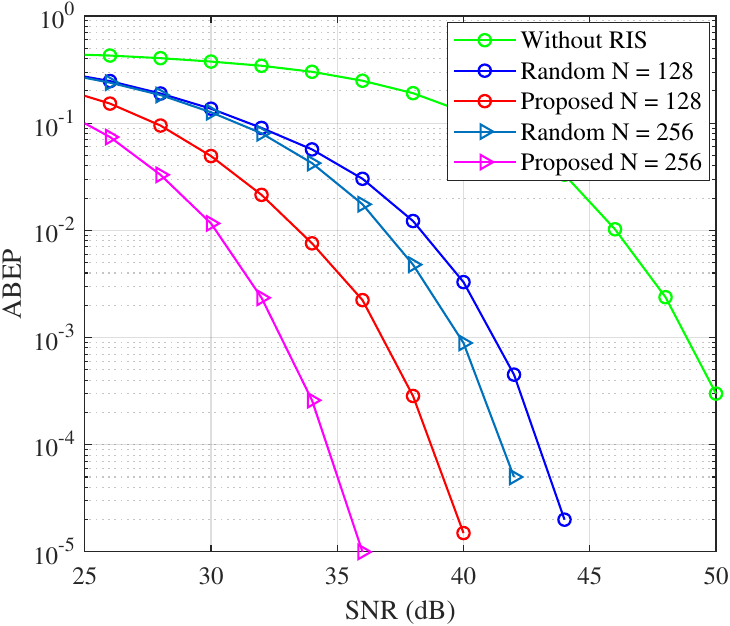}\\
  \caption{Comparison of ABEP performance of the proposed scheme with different benchmark schemes.}\label{fig1}
\end{figure}

\begin{figure}[t]
  \centering
  \includegraphics[width=4.2cm]{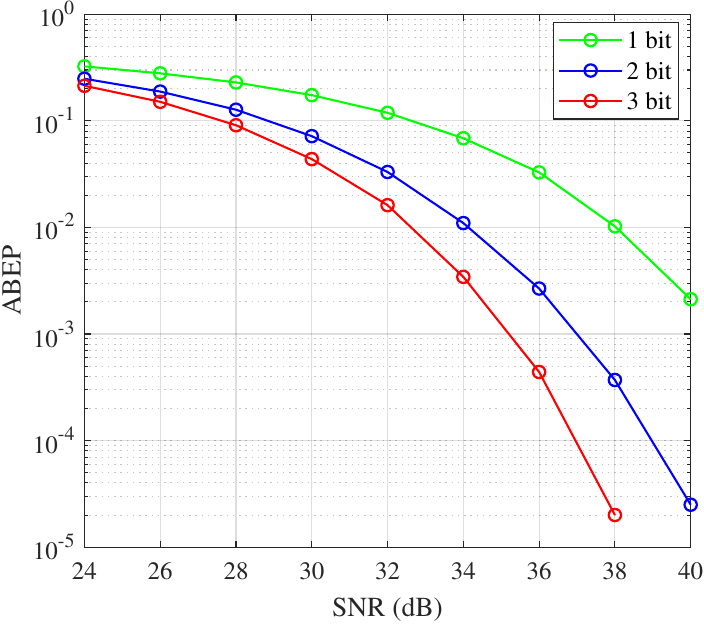}\\
  \caption{Impact of RIS phase quantisation bits on ABEP performance.}\label{fig2}
\end{figure}

\section{Simulation Results}
In this section, we evaluate the reliability performance of the proposed RIS-SSK-MIMO system with simulation results.
To enhance the received signal, we place RIS close to Rx. We consider a two-dimensional coordinate system, where the coordinates of Tx, RIS, and Rx are (0 m, 10 m), (40 m, 4 m), and (40 m, 0 m), respectively.
Besides, we uniformly set the Rician factor to $\kappa = 3$ for both Tx-RIS and RIS-Rx links.
The path loss exponents of Tx-Rx, Tx-RIS, and RIS-Rx are configured as $2.8$, $2.2$, and $2.2$, respectively.

Fig. \ref{fig1} shows the ABEP performance of the proposed RIS-SSK-MIMO scheme compared to other benchmarks when the number of transmit and receive antennas are both 4 and the discrete phase shift of RIS is 3 bits.
From Fig. \ref{fig1}, we can observe that the higher the number of RIS elements, the better the ABEP performance of the RIS-SSK-MIMO scheme, which is due to the fact that the larger the value of $L$, the stronger the reflecting energy arriving at the Rx beam.
Moreover, we also compare the RIS random reflection case. Undoubtedly, the ABEP performance obtained from the RIS random reflection phase-shift is worse than the proposed scheme, which is because the random reflection only increases the number of paths to reach the Rx and does not take full advantage of the reflected beam adjustment of the RIS.
In addition, we also compared the case without RIS assistance. The ABEP performance obtained in this case is found to be the worst, which highlights the important advantage of RIS.

In Fig. \ref{fig2}, we present the impact of the number of discrete phase shift bits of the RIS on the ABEP performance of the proposed RIS-SSK-MIMO system, where the number of equipped antenna elements is 4 for both Tx and Rx, and the number of reflecting units of the RIS is 128.
As expected, the larger the number of quantization bits of the RIS reflection phase shift, the better the ABEP performance of the proposed RIS-SSK-MIMO system becomes. This is because the error between the RIS phase shift and the continuous phase shift is smaller.
According to \cite{zhu2024ris}, we observe that the phase shift quantization bit of RIS is 3, the gap between the ABEP curve and that with continuous phase shift can be neglect. To this end, we plot the quantization bits of 1, 2, and 3 bits in Fig. \ref{fig2}.

\begin{figure}[t]
  \centering
  \includegraphics[width=4.2cm]{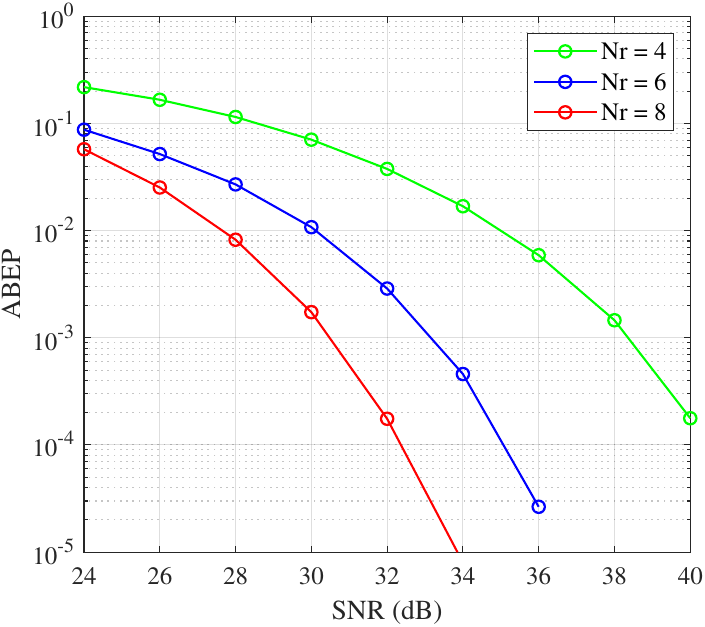}\\
  \caption{Impact of the number of receive antennas on ABEP performance.}\label{fig3}
\end{figure}

In Fig. \ref{fig3}, we plot the ABEP versus SNR for the proposed RIS-SSK-MIMO system with different numbers of receive antennas, where the number of transmit and receive antenna elements is 4, the phase quantisation bit of RIS is set to $Q=3$, and the number of RIS units is $L=128$.
As can be seen from Fig. \ref{fig3}, the system ABEP significantly decreases as the number of receive antennas increases from $N_r=4$ to $N_r=8$. This phenomenon can be explained by the fact that increasing the number of receive antennas increases the received signal energy. Moreover, increasing the dimensionality of the receive antennas provides more freedom in the SSK signal design and makes the equivalent channel responses more different from each other, i.e., the gap between the adjacent smallest constellation points of the constellation diagram becomes larger.

\begin{figure}[t]
  \centering
  \includegraphics[width=4.2cm]{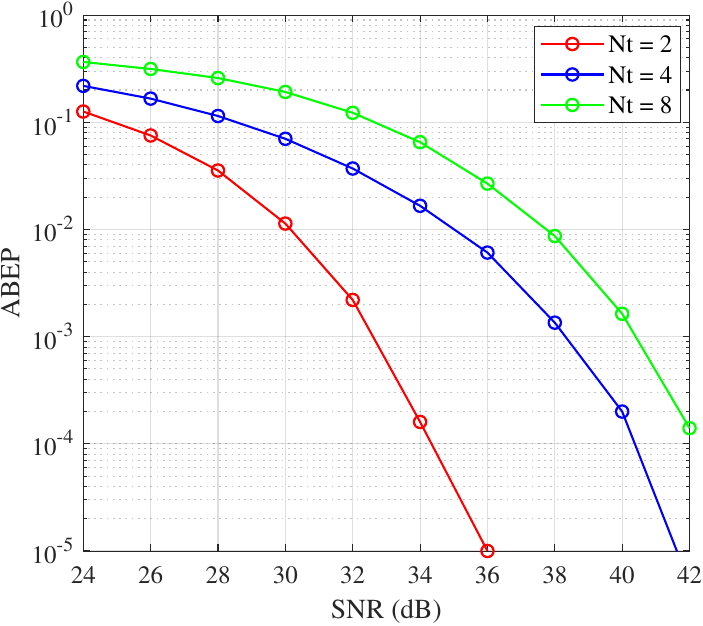}\\
  \caption{Impact of modulation order on ABEP performance.}\label{fig4}
\end{figure}

Fig. \ref{fig4} illustrates the ABEP performance of the proposed RIS-SSK-MIMO system with respect to the number of transmit antennas $N_t$, where the number of receive antenna elements, the phase quantisation bits of the RIS, the number of RIS units are configured to be $N_r = 4$, $Q=3$, and $L=256$, respectively.
It is observed that the ABPE performance of the proposed RIS-SSK-MIMO system deteriorates as the number of transmit antennas $N_t$ increases. This is because the modulation order of SSK is related to the number of antennas, and a larger number of antennas corresponds to a higher modulation order, which makes it difficult to decode the transmit antenna.


\section{Conclusion}
In this paper, a discrete RIS-assisted SSK-MIMO scheme is studied.
To improve the ABEP performance of the proposed scheme, we formulate a optimization problem to maximize the minimum Euclidean distance of constellation points under the discrete reflecting beamforming constraint. In addition, a reflecting beamforming optimization method based on a penalty alternating algorithm is proposed and the convergence of the algorithm is proved. Simulation results show that the performance of the proposed RIS-SSK-MIMO system is better than the benchmark and indicate that factors such as RIS elements, phase quantization bits, and the number of antennas affect the ABEP performance.

\end{document}